\documentclass[]{interact}

\usepackage{longtable}
\usepackage{amsmath}
\usepackage{booktabs}
\usepackage{float}
\usepackage{booktabs}
\usepackage{multirow}
\usepackage[]{geometry}
\usepackage{changepage}
\usepackage{graphicx,colortbl,textcomp,wrapfig}
\usepackage{fancyhdr,setspace,rotating}
\usepackage{ltxtable,tabularx,booktabs,multirow,dcolumn}
\usepackage{amssymb}% http://ctan.org/pkg/amssymb
\usepackage{pifont}% http://ctan.org/pkg/pifont
\newcommand{\cmark}{\ding{51}}%
\usepackage[table,xcdraw]{xcolor}
\usepackage{multicol}
\usepackage{epstopdf}
\usepackage[caption=false]{subfig}
\usepackage[numbers]{natbib}
\bibpunct[, ]{(}{)}{;}{a}{,}{,}

\newcommand{\be}{\begin{equation}}
\newcommand{\ee}{\end{equation}}
\newcommand{\bea}{\begin{eqnarray}}
\newcommand{\eea}{\end{eqnarray}}

\renewcommand\bibfont{\fontsize{10}{12}\selectfont}

\graphicspath{{.}}

\begin{document}

%\articletype{International Journal of Logistics Research and Applications}

\title{Developing a Resilient, Robust and Efficient Supply Network in Africa}
%\maketitle    %This hides authors for double blind

\author{
\name{Bruce A. Cox\thanks{CONTACT Bruce A. Cox. Email: BruceACox1@gmail.com}$^1$}
\name{Christopher M. Smith$^1$}
\name{Timothy W. Breitbach$^1$}
\name{Jade F. Baker$^1$}
\name{Paul P. Rebeiz$^2$}
\affil{[1] Air Force Institute of Technology, 2950 Hobson Way, Wright-Patterson AFB, OH 45433, USA}
\affil{[2] The Perduco Group, 2647 Commons Blvd, Beavercreek, OH 45431, USA}
}

\maketitle %This displays authors

\begin{abstract}
Supply chains need to balance competing objectives; in addition to efficiency they need to be resilient to adversarial and environmental interference, and robust to uncertainties in long term demand.  Significant research has been conducted designing efficient supply chains, and recent research has focused on resilient supply chain design.  However, the integration of resilient and robust supply chain design is less well studied. This paper develops a method to include resilience and robustness into supply chain design. Using the region of West Africa, which is plagued with persisting logistical issues, we develop a regional risk assessment framework, then apply categorical risk to the countries of West Africa using publicly available data. Next, we develop a mathematical model leveraging this framework to design a resilient supply network that minimizes cost while ensuring the network functions following a disruption. Finally, we examine the network's robustness to demand uncertainty via several plausible emergency scenarios.
\end{abstract}  


\section{Introduction}
%Lead in with something broader than just the US. OR, just jump straight to next paragraph. Motivate it with global attention on Africa--maybe overseas development dollars and foreign government money coming in. The United States Africa Command (USAFRICOM) is one of the ten unified combatant commands of the United States armed forces.   AFRICOM was established in 2007 in response to the 2001 World Trade Center terrorist attacks, after which the public viewed underdeveloped countries as a threat to national security \citep{call2016lingering}\citep{pham2014development}\citep{van2009us}.   The goal of AFRICOM is to advance U.S.  interests relating to Africa by increasing its security capabilities and ability to respond to catastrophes and international threats \citep{BCA2017}.

The Ebola outbreak and the rise of terrorist organizations such as Boko Haram and the Islamic State in Africa have given Western Africa a recent place of prominence on the global stage. In conjunction with this rise in global awareness, multiple agencies across the U.S. Government--from the Center for Disease Control to the Department of State and the Department of Defense, have increased their presence in the region. Western Africa, encompassing 18 countries roughly the size of the continental United States, has become the region in Africa of highest logistical concern for the U.S. Government. With this increased presence comes a critical need for reliable logistics support.

A key challenge facing the US, the international community and commercial companies establishing operations in Africa--especially Western Africa given the number of countries and the infrastructure condition, is the development of a robust, resilient and efficient supply chain network. Historically, the US Government supported humanitarian, military or other operations through the use of ad hoc and contingency-scheduled missions. However, this has become costly and unreliable as the required support in the region has increased. The absence of an efficient distribution systems hinders the U.S. and international community's ability to respond to humanitarian crises and help local governments combat terrorism. 

There have been recent efforts to establish such a network in West Africa, and we examine a hub and spoke distribution network in which material originating in the U.S. or Europe flows into a main hub on the West African coast. In addition to the main hub, there are six transhipment nodes spread throughout the region that support up to 53 additional potential demand locations. It is a multi-modal distribution network in which product can be transported by air, ground or water. This network was established to increase both the the efficiency and reliability of support for U.S. Government activity in Western Africa. However, the network's resilience has not been assessed, and this is of particular concern given the challenges currently facing the region. 

The research goal for this paper is to develop a resilient, robust and efficient supply chain network for West Africa. This is achieved via the following contributions.  First, we propose a risk analysis framework to codify the risk of supply chain disruptions associated with the region's environmental challenges, natural barriers, and internal conflicts.  Second, we propose optimal design changes to the West Africa supply chain to ensure the resulting network is resilient to disruption by the identified risks.  Specifically, we propose optimal primary and backup routes for material flow through the network.  Finally, we examine changes to the set of transshipment nodes, which induces changes to primary and backup routes, to ensure the network's robustness under historically plausible stochastic demand scenarios.  

The paper is structured as follows. Section 2 presents a literature review of types of networks designs, along with codifying meanings for \textit{disruptions, resilience} and \textit{robustness} in the context of graph theory.  Section 3 outlines the risk analysis framework, and the underlying mathematical model parameters, inputs and formulation.  Section 4 outlines the disruption scenarios and model variations considered, then analyzes several model instantiations.  Section 5 summarizes research findings and recommendations.  Section 6 concludes by summarizing research and providing future research directions.

\section{Resilient, Robust and Efficient Network Designs}
Developing a supply chain for West Africa is a Supply Chain Network (SCN) design problem, a subset of the facility location problem. (See, for instance, \citep{mirchandani1990discrete, daskin2011network} for excellent treatments of the facility location problem.)  Specifically, the SCN design problem models the supply chain as a network via graph theory \citep{wagner2010assessing}, with nodes in the graph representing the locations of interest (i.e., suppliers, warehouses, and or customers) and the graph's arcs representing the transportation routes between these locations.  Numerous types of supply networks have been used across the Department of Defense and industry in order to balance resilience, cost and effectiveness. We examine the pros and cons of three of these categories: ad hoc, efficient, and resilient.

Ad hoc, or Just-in-Time (JIT), networks create nodes based on customer demand and network necessity. These networks enable fast shipments and are reactive to changes in customer demand, making them useful in disaster-prone situations. Ad hoc networks are regularly utilized in supply chains, as well as in healthcare, computer, and cellular network designs. As Golhar and Stamm \citep{golhar1991just} point out, the basic practice of ad hoc networks is based on four major tenets. Those tenets are: employee input in decision making, waste elimination, supplier participation, and quality control. Employee input initiates the creation of a node. Because a node is created per customer demand, resources needed to maintain that node are unnecessary until the node’s implementation, i.e., waste is eliminated. Finally, supplier participation ensures a quick response to customer demand, thus, increasing customer satisfaction and quality control \citep{kannan2005just}. 

A variant of JIT networks is the dynamic JIT network. Under this class of networks, demand can change over time, between locations, and at different levels of the network \citep{lai2003study}. \citet{marina2001demand} characterize ad hoc networks as having mobile hubs that develop and maintain routes as needed. They create distance vectors for dynamic, on-demand networks using multiple path routing, leading to faster and more efficient networks. \citet{alkhuder2018securing} completed a study on wireless sensory networks. His article presents methods that assist a node’s instantaneous decision making or ad hoc function.  JIT networks are reactive, fast and dynamic. Consequently, ad hoc networks tend to be useful in disaster situations; they can still perform and send communication in a relatively quick amount of time when unexpected mishaps occur \citep{fujiwara2005ad}. The main disadvantage of ad hoc networks comes from their non-scalability – as the network grows, oversight of employees becomes increasingly difficult. Further, some JIT applications can be more expensive than those that permanently position nodes and routes. 

Another network class is efficient networks, whose goal is to obtain the most benefit from a network while using the fewest number of permanent locations. Efficient networks are commonly used in warehouse assignment and location problems, in which warehouses are optimally placed to minimize the total number of warehouses and to possibly reduce transportation costs.  \citet{jackson1996strategic} showed that efficient networks are either complete graphs, star graphs, or graphs with no links. Further,  \citet{heydari2015efficient} showed that any two nodes in an efficient network have a maximum distance of two connections between them. If this is not the case, the routes can be restructured to obtain a network that achieves similar or better results. In the realm of applications, \citet{jacyna2017multi} implemented a genetic algorithm to efficiently position warehouses based on a multicriteria logistical network. Their algorithm generated a network with efficiently placed nodes that had the same performance as networks with more nodes. 

In network design, hub-and-spoke networks are a type of efficient networks. Under a hub-and-spoke structure, demand locations (spokes) are all connected to warehouses (hubs).  \citet{yang2017hub} discuss the efficiency of this type of design in which hubs are situated to minimize the connection length between spokes. This corresponds to placing hubs in a centralized location, which, in turn, minimizes the number of hubs. As such, one can conclude that the warehouse location problem and the hub-and-spoke design problems both aim to achieve network efficiency. Although efficient networks provide great benefit by reducing the number of components while simultaneously satisfying demand, such networks pose notable challenges.  

\citet{jackson1996strategic} note that efficient networks are not always stable.  Jackson and Wolinksy describe stability as a network’s ability to alter its connections without damaging its function (i.e., resilient to disruption). However, in an efficient network, removal of a link or of a node and its links could have serious ramifications, potentially sabotaging any productivity gained by using an efficient network. This motivates the consideration of the third common type of network, a resilient network.  \citet{mensah2014developing} and \citet{sheffi2005supply} define resilience as the network’s capacity to return to its original state after disruption or to mitigate disruption of network flow entirely.  Hutchison and Sterbenz define resilience as a network's ability to maintain its service level during disruptions \citep{hutchison2018architecture}.  However, both of these definitions fail to explicitly define what is meant by a 'disruption'.  Indeed \citet{kim2015supply} note this exact problem and provide sixteen different academic definitions for a disruption!  For the purposes of this research we utilize a graph-theoretic definition for resilience similar to that used by \citet{kim2015supply}.  Specifically we state that a disruption is any action (natural or man made) that removes an arc or a node from the network.  This is a rather binary definition, as of course in the real world partial degradation is more likely, but it establishes a conservative design. 

\citet{jackson1996strategic} exemplify a tetrahedron as a stable network, as all links are connected to one another.  Thus even with removal of a arc or node the network preserves connectivity between the nodes.  Thus, networks can become more resilient by adding nodes or more arcs.  \citet{zhalechian2018hub} argue that adding resilience to their hub-and-spoke model could be achieved by connecting each node to multiple hubs.  They construct a fortified hub-and-spoke network, which resulted in a resilient network.  This network has built-in proactive and reactive capabilities, protecting it from natural disasters, terrorist attacks, and internal network problems.  In another paper, \citet{ng2018transport} address network risk management by classifying nodes and arcs based on their geographic characteristics.  Using this classification, Ng et al. were able to mitigate the risk of network delays by assessing each node's ease of access as well as its suitability as a hub or spoke.   \citet{sadghiani2015retail} use a set covering approach to ensure demand locations are covered by multiple suppliers based on characteristics of the demand locations in order to increase resilience.  Such approaches focus on ensuring SCN resilience by adding routes and nodes that preserve all locations' ability to meet their demand.  These network additions are targeted at the location level.  Such approaches are broadly aligned with the technique we present in section \ref{FrameworkAndModel}.

Network robustness is an additional topic relevant to SCN design, but it is often confounded with resilience.  \citet{yang2017hub} point out that designing a network corresponds to a long-term strategic decision under a dynamic and changing environment and that due to variations in the network’s geospatial inputs, a robust network could be required rather than an efficient network. \citet{klibi2010design} define robustness as ‘‘the quality of a SCN to remain effective for all plausible futures.’’  In general robustness is a network's ability to function 'well' under a diverse set of potential future scenarios \citep{snyder2006planning, dong2006development, ben2009robust, mulvey1995robust}.  The topic of SCN robustness is still quite nascent, and to the best of our knowledge the only other paper presenting both a resilient and scenario-based robust optimization SCN design is  \citep{sadghiani2015retail}.   

The supply chain proposed to serve demand in Western Africa is an efficient network.  However, a focus on efficiency can inhibit a network's ability to function in emergencies or disaster situations (lack of resilience), as well as hinder its ability to function well under future scenarios with changed system parameters (lack of robustness).  The ability to operate under emergency and crisis situations is paramount in humanitarian and military operations. The need for this logistics network will remain far into the future. Our study adds elements to ensure both resilience and robustness are considered in the network supporting West Africa.  This research develops a framework to identify and quantify risk in West Africa, then develops and optimizes a mathematical model to find the best routes to ensure necessary network resilience, and finally uses stochastic analysis of plausible demand scenarios to ensure network robustness.

\section{Risk Analysis and Model Formulation}\label{FrameworkAndModel}
\par To address the logistical challenges present in West Africa--large distances, poor infrastructure, insecure routes, etc., the supply chain  was restructured to achieve a higher service level and increased efficiency.  The network operates under a hub-and-spoke design with a main hub, 6 transshipment nodes and 53 other potential demand locations.   Note that any of the transshipment nodes may also experience a demand.

\par Table \ref{tab:demandlocations} shows the 18 West African countries that are included in the analysis, along with the number of potential demand nodes within each country.  The European hub serves as a “dummy node”, where supplies originate from this location and ship to the main hub.  The main hub of entry into West Africa, and the other six transshipment locations, were determined by decision makers prior to the undertaking of this study. We seek to assess and increase the network's resilience given a set of decisions that were made for reasons outside the scope of this analysis.

\begin{table}[]
    \centering
    \begin{tabular}{|c|c|}
        \hline 
        \textbf{Country} & \textbf{Number of Nodes}  \\ \hline 
         Benin          &   1 \\ \hline 
         Burkina Faso*  &   8 \\ \hline 
         Cameroon*      &   5 \\ \hline 
         Chad           &   4 \\ \hline 
         Ivory Coast    &   1 \\ \hline 
         Gabon          &   1 \\ \hline 
         Ghana          &   3 \\ \hline 
         Guinea         &   1 \\ \hline 
         Liberia        &   1 \\ \hline 
         Mali           &   7 \\ \hline 
         Mauritania     &   4 \\ \hline 
         Niger          &   3 \\ \hline 
         C. African Rep.  &   1 \\ \hline 
         Niger*         &   7 \\ \hline 
         Nigeria        &   8 \\ \hline 
         Senegal*       &   3 \\ \hline 
         Sierra Leone   &   1 \\ \hline
         Togo           &   1 \\ \hline
    \end{tabular}
    \caption{Supply Chain consists of 60 total demand locations plus origin node in Europe.  Countries identified as having transshipment nodes in original architecture are indicated with a '*'.}
    \label{tab:demandlocations}
\end{table}

\begin{figure}[h]
    \centering
    \includegraphics[width=0.95\textwidth]{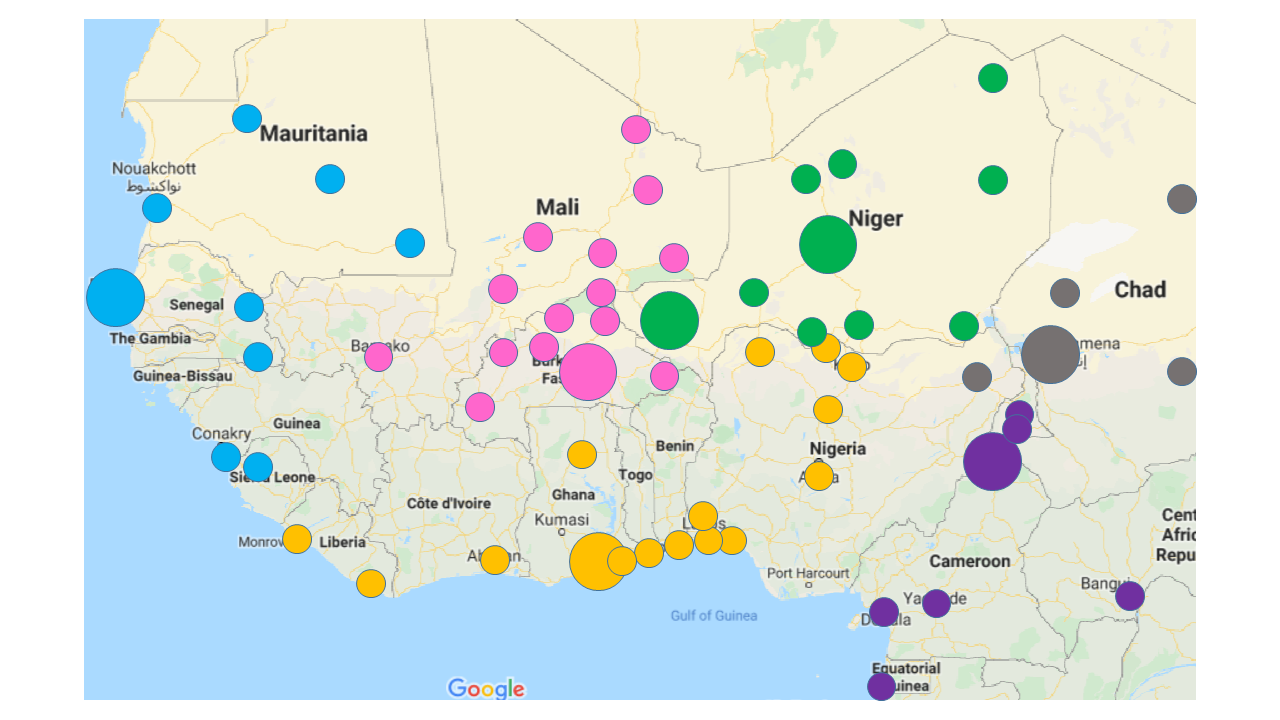}
    \caption{Hub and spoke map of the West Africa Logistics Network.  The seven large nodes are all hubs (i.e transshipment nodes).  The small nodes are spokes.  The spokes are color coded to match their primary hub.}
    \label{tiermap}
\end{figure}

Figure \ref{tiermap} displays a map of the region, in which the large nodes (hubs) are transshipment facility locations and the small nodes (spokes) are demand locations.

\subsection{Risk Framework}\label{RiskFramework}
The network configuration as first established does not implement any backup routes for material flow, possibly leaving nodes inaccessible should a hub, or route, disruption occur.  Granted, air routes are flexible by nature, but the long distances between locations coupled with the delays inherent to country clearance and border issues, necessitates a relatively fixed delivery schedule.  The primary goal of this research is to identify an optimal variant of the network which adds resiliency to disruption while maintaining efficiency.  Towards that end our model requires each location to be connected to more than one transshipment node.  However, determining 'appropriate' levels of redundancy is non-trivial as these decisions necessitate a risk analysis framework and implementation--our first major contribution.  We focus our risk analysis on three areas of interest for every country: internal conflict, infrastructure, and terrain/climate.  

Table \ref{tab:resiliency framework} gives a short description of these three categories, along with the available scores, and the meaning of each score by category.  Classifications for conflict are determined by the U.S. Department of State travel advisory, and classifications for infrastructure are determined by the Central Intelligence Agency World Fact Book \citep{ciafactbook}\citep{consular2018africa}. Classification for Terrain/Climate was accomplished by the research team, thus presenting one subjective view. However, the classifications could be adjusted based on subject matter expertise or changing contextual conditions.  Our unique approach incorporates the risk framework into the supply chain network design. 

\begin{table}[h]
\centering
\caption{Resiliency risk analysis framework: Description and scores for three considered categories of internal conflict, infrastructure, and terrain/climate. }
\begin{tabular}{|c|c|l|l|l|}
\hline
\multicolumn{2}{|l|}{}            & \multicolumn{1}{c|}{Conflict}                                                               & \multicolumn{1}{c|}{Infrastructure}                                                 & \multicolumn{1}{c|}{Terrain/Climate}                                                 \\ \hline
\multicolumn{2}{|c|}{Description} & \begin{tabular}[c]{@{}l@{}}Internal disputes and drug \\ and human trafficking\end{tabular} & \begin{tabular}[c]{@{}l@{}}Air, land, \\ and sea capabilities\end{tabular}          & \begin{tabular}[c]{@{}l@{}}Disruptions relating to \\ natural phenomena\end{tabular} \\ \hline
\multirow{3}{*}{Score}     & 0    & \begin{tabular}[c]{@{}l@{}}Exercise normal precaution\\  when traveling\end{tabular}        & \begin{tabular}[c]{@{}l@{}}Ranked  in Top 3rd in \\ World Fact Book\end{tabular}    & No Issues                                                                            \\ \cline{2-5} 
                           & 1    & \begin{tabular}[c]{@{}l@{}}Exercise increased caution \\ when traveling\end{tabular}        & \begin{tabular}[c]{@{}l@{}}Ranked  in Middle 3rd in \\ World Fact Book\end{tabular} & Mild Issues                                                                          \\ \cline{2-5} 
                           & 2    & \begin{tabular}[c]{@{}l@{}}Do not travel/reconsider \\ travel\end{tabular}                  & \begin{tabular}[c]{@{}l@{}}Ranked  in bottom 3rd in\\  World Fact Book\end{tabular} & Severe Issues                                                                        \\ \hline
\end{tabular}
\label{tab:resiliency framework}
\end{table}

Table \ref{tab:resiliency analysis by country} presents the results for each country.  In order to determine an aggregated risk for each country we place equal weight on each of three criteria and sum the individual scores to assess an overall risk score for each country. It should be noted that this framework does not preclude the application of non-equal weights if a subject matter expert determines one of the three categories to be more, or less, significant then the others.  

Lastly, we compute a required resiliency level, ${R_c}$, for each country $c$ based on the aggregated risk score for each country.  Each location is then assigned a required resiliency level equal to its countries level. i.e., ${R_i}$=${R_c}$. For the purposes of this research the following step function was used to assign required resilience levels based on the aggregated risk. 
\begin{equation}
R_i= 
  \begin{cases}%{llr}
    2   & \text{if Aggregated Risk = 0 or 1} \\
    3   & \text{if Aggregated Risk = 2 or 3} \\
    4   & \text{if Aggregated Risk = 4} 
  \end{cases}
\end{equation}

\begin{table}[h!]
\centering
\caption{Resiliency risk analysis by country.  Explicit risk scores are first assigned to each country by category, score are then aggregated together, and finally a resiliency metric $R_c$ is assigned to each country based on it's aggregated risk.}
\begin{tabular}{|c|c|c|c|c||c|c|}
\hline
\multirow{2}{*}{Country}                                             & \multirow{2}{*}{Conflict} & \multirow{2}{*}{\begin{tabular}{l}Infra-\\structure\end{tabular}} & \multicolumn{2}{c|}{Terrain/Climate}                                                                          & \multirow{2}{*}{\begin{tabular}{l} Aggregated\\ Risk \end{tabular}} & \multirow{2}{*}{$R_c$}    \\ \cline{4-5}
                                                                     &                           &                                 & Description                                                                                           & Score &                            &                        \\ \hline
Benin                                                                & 0                         & 2                               &                                                                                                       & 0     & 2                          & 3                      \\ \hline
\begin{tabular}[c]{@{}c@{}}Burkina\\   Faso\end{tabular}             & 2                         & 1                               &                                                                                                       & 0     & 3                          & 3                      \\ \hline
Cameroon                                                             & 1                         & 1                               & \begin{tabular}[c]{@{}c@{}}Volcano in \\ W. Africa    with\\ fatal levels of gas\end{tabular} & 2     & 4                          & 4                      \\ \hline
\begin{tabular}[c]{@{}c@{}}Central\\African\\ Republic\end{tabular} & 2                         & 1                               &                                                                                                       & 0     & 3                          & 3                      \\ \hline
Chad                                                                 & 2                         & 1                               &                                                                                                       & 0     & 3                          & 3                      \\ \hline
Gabon                                                                & 0                         & 1                               &                                                                                                       & 0     & 1                          & 2                      \\ \hline
Ghana                                                                & 0                         & 0                               &                                                                                                       & 0     & 0                          & 2                      \\ \hline
Guinea                                                               & 1                         & 1                               & \begin{tabular}[c]{@{}c@{}}Monsoonal rainy \\ season\\ (June to November)\end{tabular}      & 2     & 4                          & 4 \\ \hline
\begin{tabular}[c]{@{}c@{}}Guinea\\   Bissau\end{tabular}            & 2                         & 2                               &                                                                                                       & 0     & 4                          & 4 \\ \hline
Ivory Cost                                                           & 1                         & 0                               & \begin{tabular}[c]{@{}c@{}}Possible torrential\\ flooding during\\ rainy season\end{tabular}          & 1     & 2                          & 3 \\ \hline
Liberia                                                              & 0                         & 0                               &                                                                                                       & 0     & 0                          & 2 \\ \hline
Mali                                                                 & 2                         & 1                               &                                                                                                       & 0     & 3                          & 3                      \\ \hline
Mauritania                                                           & 2                         & 1                               &                                                                                                       & 0     & 3                          & 3                      \\ \hline
Niger                                                                & 2                         & 1                               &                                                                                                       & 0     & 3                          & 3                      \\ \hline
Nigeria                                                              & 2                         & 0                               &                                                                                                       & 0     & 2                          & 3                      \\ \hline
Senegal                                                              & 0                         & 1                               &                                                                                                       & 0     & 1                          & 2                      \\ \hline
\begin{tabular}[c]{@{}c@{}}Sierra\\   Leone\end{tabular}             & 1                         & 1                               & \begin{tabular}[c]{@{}c@{}}Rainfall makes it \\ one of the wettest\\ W. African coasts\end{tabular}    & 1                     & 3                   & 3                      \\ \hline
Togo                                                                 & 1                         & 1                               &                                                                                                       & 0     & 2                          & 3                      \\ \hline
\end{tabular}
\label{tab:resiliency analysis by country}
\end{table}

\subsection{Mathematical Formulation}\label{MathematicalFormulation}
We developed an explicit mathematical formulation of the distribution network in West Africa as a mixed integer linear program in order to determine the optimal location for transshipment locations and route connections.  To provide a precise statement of this problem, we define:
\\ \\
Sets:
\\
\textit{I}: \text{Set of locations}, \textit{i} $\in$ \textit{I}=\text{\{1, $\dots$, 61\}}.
\\
\textit{K}: \text{Set of transportation types}, \textit{k} in \textit{K}=\text{\{1=air, 2=land, 3=sea\}}.
\\
\\
Parameters:
\\
\textit{C} =	\text{Maximum number of spokes a transshipment node can support.}
\\
\textit{${D_{i,j}^k}$} = Distance between locations \textit{i} and \textit{j} using transportation type \textit{k}.
\\
\textit{${M_i}$} = Amount of material demanded at location \textit{j}.
\\
\textit{${R_j}$} = Number of connections that location \textit{j} is required to have with warehouses.
\\
\textit{S} = Amount of supply material entering the main hub.
\\
\textit{W} = Number of warehouses required to open.
\\
\\
Decision Variables:
\\
${x_i}$	= Binary variable; equals 1 if warehouse \textit{i} is open, 0 else.	
\\
${y_{i,j}^{k}}$ = Binary variable; equals 1 if a route is established between locations \textit{i} and \textit{j} for transport type \textit{k}, 0 else.
\\
${z_{i,j}^k}$	= Amount of material shipped from location \textit{i} to \textit{j} via transportation type \textit{k}.\\

The network allows material to be delivered using three methods of transportation: air, land, and sea.  Our underlying network is assumed fully connected for all three transportation types, with distances calculated as follows:  for any two locations \textit{i},\textit{j} $\in$ \textit{I}, we compute the actual distance ${D_{i,j}^k}$ for transportation type \textit{k}.  If a transportation type is infeasible between two locations, that specified route receives an input of one million miles, ensuring  the model does not select an infeasible route.  Since a core tenant of the network is a single point of entry from Europe, the initial version of the model established as infeasible all other routes to West Africa.  

Distance by air was calculated with the Euclidean distance formula, using each location's latitude and longitude coordinates, as determined by Google Earth.  Only pairs of locations with airports and runways within 20 miles of each endpoint had calculated air distances.  Shared airports between two locations were precluded.  Ownership of an airport was determined by closest distance.

Ground transport distances between locations were estimated using the recommended path from Google Maps and were only assigned if the route linking the pair of locations could be traversed by land vehicles.  Specifically, land transportation was deemed unsuitable if another form of transportation was also required, e.g., both ferry and ground transportation.  

Finally, sea transportation distances were calculated using the website, Sea-Distance.org, which provides nautical miles between ports \citep{seadistances}.  For consistency, the calculated nautical miles were then converted to miles.  Sea distances were only computed for pairs of locations with sea ports at each endpoint.   We assume that there is no capacity limit for each mode of transportation.

The problem of warehouse selection, network connections establishment, and inventory distribution is formulated by the following mixed integer linear program which we call the African Supply Chain Network Problem (ASCNP).

{\allowdisplaybreaks
\begin{align}
     \label{Con:ObjFun} \min_{\boldsymbol{x,y,z}} & w_1 f_1(\boldsymbol{z})+w_2 f_2(\boldsymbol{y})  \\
     \label{Con:ObjFun1} \text{s.t. } & f_1(\boldsymbol{z})= \sum_{k \in K}\sum_{i \in I}\sum_{j \in I}{D_{i,j}^k}{z_{i,j}^k},  \\
     \label{Con:ObjFun2}  & f_2(\boldsymbol{y})= \sum_{k \in K}\sum_{i \in I}\sum_{j \in I}{D_{i,j}^k}{y_{i,j}^k},  \\
     \label{Con:NumHubs} & \sum_{i \in I/{1}}{x_i} = 7,\\
     \label{Con:MeetResiliency} & \sum_{k \in K}\sum_{i \in I/{1}}{y_{i,j}^{k}} >= {R_j} \qquad \forall \textit{j} \in \textit{I}, \\
     \label{Con:MaxSpokes} & \sum_{k \in K}\sum_{j \in I}{y_{i,j}^k} <= C{x_i} \qquad \forall \textit{i} \in \textit{I},\\
     \label{Con:MaxSupply} & \sum_{k \in K}{z_{1,{22}}^k} <= S,\\
     \label{Con:RouteCapacity} & \sum_{k \in K}{z_{i,j}^k} <= S{y_{i,j}}  \qquad \forall \textit{i}, \textit{j} \in \textit{I}, \\
     \label{Con:MeetDemand} & \sum_{j \in I}\sum_{k \in K}{z_{j,i}^k} - \sum_{j \in I}\sum_{k \in K}{z_{i,j}^k} >= {M_i} \qquad \forall \textit{i} \in \textit{I},\\
     \label{Con:StartingHubs} & {x_7} = 1,\hspace{.1cm} {x_{12}} = 1,\hspace{.1cm} {x_{19}} = 1,\hspace{.1cm} {x_{22}} = 1,\hspace{.1cm} {x_{38}} = 1,\hspace{.1cm} {x_{46}} = 1,\hspace{.1cm} {x_{58}} = 1,\\
     \label{Con:StartingRoute} & {y_{1,{22}}^1} = 1,\hspace{.1cm} {y_{1,{22}}^2} = 0,\hspace{.1cm} {y_{1,{22}}^3} = 1\\
     \label{Con:Binary} & {x_i} \in \{0,1\},\hspace{.1cm} {y_{i,j}^k} \in \{0,1\},\hspace{.1cm} {z_{i,j}^k} \in \mathbb{Z} \qquad \forall \textit{i}, \textit{j} \in \textit{I},\hspace{.1cm} \forall \textit{k} \in \textit{K}.
\end{align}}

ASCNP is structured as a weighted sum multi-objective optimization problem, other multi-objective approaches such as the $\epsilon$-constraint approach and the lexicographic method \citep{ehrgott2005multicriteria}, were considered but not deemed appropriate.  The first term in the objective function (\ref{Con:ObjFun1}) minimizes the total distance of shipped items.  This can be thought of as a proxy for minimizing the shipping costs for the primary routes in the network.  While distance is an imperfect metric, attributing accurate costs for land, air and sea shipments was beyond scope of this research, so material transit distance was used as a cost proxy.  The second term (\ref{Con:ObjFun2}) minimizes the distance of established routes.  This term is necessary to ensure that the backup routes established for resiliency are not chosen arbitrarily.  Constraint (\ref{Con:NumHubs}) requires the model to have a total of seven transshipment nodes.  In context of our base model, these locations are already predetermined as seen in table \ref{tab:demandlocations}, however variants of the ASCNP enable exploration of more or less then these seven base hubs.  Constraint (\ref{Con:MeetResiliency}) ensures each destination location \textit{j} $\in$ \textit{I}  is serviced by at least ${R_j}$ transshipment nodes.  The number of transshipment node connections, ${R_j}$, required to ensure network resilience commensurate with location risk was determined according to the methodology in section \ref{RiskFramework}. Constraint (\ref{Con:MaxSpokes}) imposes an upper bound, \textit{C}, on the number of outbound connections a transshipment node can support (${x_i}$=1) and enforces that no connection ${y_{i,j}}$ can be established between locations \textit{i} and \textit{j} unless \textit{i} is a transshipment node.  Our initial assumption is that all transshipment nodes can support the same number of locations, and further that all transshipment nodes can support the full set of locations, thus \textit{C} = 60 for our base case.   Constraint (\ref{Con:MaxSupply}) imposes an upper bound on the sum of flow of material into the region and ensures it is no greater than the supply, \textit{S}, provided by the source location.  Constraint (\ref{Con:RouteCapacity}) ensures that no material can flow between locations \textit{i} and \textit{j} unless a connection is established between them (${y_{i,j}}$=1).  Constraint (\ref{Con:MeetDemand}) represents the flow constraint at location $i$ and ensures that the net flow of material to a location meets it's demand.  (i.e., Total flow of material delivered to a location minus total flow of material shipped out from that location is at least equal to its demand, ${M_i}$.)  As a result, locations can accumulate excess material, an occurrence not atypical of real-world instances of warehouses having a surplus of material to insulate network from demand spikes.  Constraint (\ref{Con:StartingHubs}) sets the seven transshipment locations to those determined by when the original network was established, while constraint (\ref{Con:StartingRoute}) establishes a main hub the point of entry to the network from Europe.  Finally, the binary and integrality constraints are enforced in Constraint (\ref{Con:Binary}).  

Given the current assumption that each arc in the network is unconstrained, (i.e., that each mode of transportation can ship an unlimited quantity of materials) the transportation type providing the shortest route between any two locations will always be selected.  As a result, we could pre-process the distance parameters by first determining the shortest distance transportation modality for all routes \textit{i}, \textit{j} $\in$ \textit{I}, then replacing the term ${D_{i,j}^k}$ in the objective function by ${D_{i,j}}$ where ${D_{i,j}} = \min_{k \in K} {D_{i,j}^k}$ and the decision variable ${z_{i,j}^k}$ by ${z_{i,j}}$.  However, the model as provided enables the realistic addition of arc capacities, and reaches optimality in seconds. 

% Table 4 shows a sample of the model's output, illustrating for each established connection, the distances and shipment quantities for air, land, and sea travel.  In each case, items are shipped between a transshipment node and a destination via the transportation mode with the shortest distance.  For example, Accra (node 22) ships all items to Cotonou (node 2) via sea as this is the shortest method of transportation, whereas Ouagadougou (node 7) ships to Dori (node 5) via land, the only feasible means of transportation between these locations.

% \begin{table}[H]
% \centering
% \caption{Shipment sample output}
% \begin{tabular}{|c|c|c|c|c|c|c|ll}
% \cline{1-7}
% \multirow{2}{*}{Connection} & \multicolumn{3}{c|}{Distance (miles)} & \multicolumn{3}{c|}{Shipment} &  &  \\ \cline{2-7}
%                             & Air          & Land    & Sea          & Air      & Land     & Sea     &  &  \\ \cline{1-7}
% (22,2)                      & 185.36       & 210     & 152          & 0        & 0        & 50      &  &  \\ \cline{1-7}
% (22,11)                     & 694.56       & 974     & 711          & 50       & 0        & 0       &  &  \\ \cline{1-7}
% (7,5)                       & 1,000,000    & 167     & 1,000,000    & 0        & 50       & 0       &  &  \\ \cline{1-7}
% (7,6)                       & 1,000,000    & 147     & 1,000,000    & 0        & 50       & 0       &  &  \\ \cline{1-7}
% \end{tabular}
% \end{table}

The resilience of the resulting network is ensured by requiring each location be serviced by a calculated number of transshipment facilities. Identifying backup routes early enables contract vehicles and contingency plans to be enacted.  Locations with higher risk of disruption due to conflict, infrastructure, or terrain have higher connection requirements. Hence, in the event of a network disruption, a previously identified and established backup route could be activated, preventing demand shortages.  

\begin{figure}[h!]
    \centering
    \includegraphics[width=0.75\textwidth]{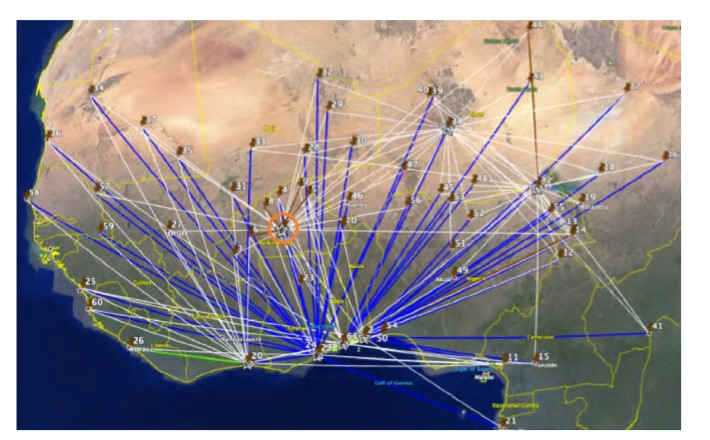}
    \caption{Default network connectivity.  White stars are transshipment hubs. Blue, brown and green lines are primary routes for air, land, and sea resp..  White lines are reserve routes.}
    \label{fig:sample connectivity output}
\end{figure}

Figure \ref{fig:sample connectivity output} displays the network output from the default ASCNP.  Since the default model fixes the transshipment hubs predetermined locations, the main output of interest is the arc map.  The shaded lines indicate the primary transportation routes for air, land, and sea respectively.  The lighter lines represent the proposed backup connections.  
%PRobably just delete--good to include one map I think but the rest can probably go. For example, the circled location is a transshipment hub (as indicated by the star) and is the primary transshipment hub for 4 locations (brown lines representing ground routes), and the backup transshipment hub for another 21 locations (white lines).

\section{Scenario-based Robustness Analysis}
The goal of this research is to design a resilient and robust network for West Africa covering 60 potential demand locations.  Resilience to potential network disruptions is handled in the ASCNP model via the $R_i$ parameter, as informed by the risk assessment framework developed in section \ref{RiskFramework}.  However, network disruptions are not the only risk that a supply network needs to be able to handle.  Scenario-based robustness analysis was used to assess how changing demand signals and/or capacity restrictions resulted in changes in the network design.  This was accomplished by analyzing different demand scenarios and network modifications.  These demand scenarios and network modifications are presented below.

\subsection{Scenarios Considered}
Due to various instability issues in West Africa, demand at different locations is prone to fluctuation.  To account for the variation in demand, we incorporated three scenarios based on their historical occurrences: epidemic, terrorist attack, and internal conflict.  For each of the scenarios, we assigned a probability of occurrence as well as the resulting percentage increase in demand to account for the heightened requirement for food, medical supplies, equipment and personnel.   It should be noted that the initial demand at each location begins at 50 units.  Table \ref{tab:crisis scenarios} summarizes each demand scenario's affected countries, likelihood of occurring, and percentage increase in demand.

\begin{table}[H]
\centering
\caption{Three crisis scenarios considered for robustness analysis along with probabilities of occurrence, locations effected, and associated increase in regional demand to handle crisis. \citep{Ebola2017africa, roby2018africa}}
\begin{tabular}{|c|c|c|c|}
\hline
            & Epidemic  & Terrorist Attack  & Internal Conflict                 \\ \hline
Likelihood  & 20\%      & 50\%              & 80\%                              \\ \hline
\begin{tabular}[c]{@{}c@{}}Countries\\   affected\end{tabular}  & \begin{tabular}[c]{@{}c@{}}Liberia, Senegal,\\ Sierra Leone \end{tabular} & Chad, Niger, Nigeria \citep{roby2018africa} & \begin{tabular}[c]{@{}c@{}}Burkina Faso,\\ Mali, Niger \end{tabular} \\ \hline
\begin{tabular}[c]{@{}c@{}}Demand\\   increase\end{tabular}    & 300\%   & 200\%    & 100\%     \\ \hline
\end{tabular}
\label{tab:crisis scenarios}
\end{table}

As multiple scenarios could occur simultaneously, there are eight possible combinations of crisis scenarios.  As determined by Table \ref{tab:crisis scenarios}, it is estimated that epidemic, terrorist attack, and internal conflict occur 20\%, 50\%, and 80\% of the time, respectively.  Therefore, using Bayes’ Theorem  we can determine the probabilities for each possible combination of crisis.  We notice that events C and BC were the most likely to occur and thus will be discussed in more depth in our model instance analyses.  The resulting eight possible scenarios are detailed in Table \ref{tab:crisis combinations}.

\begin{table}[H]
\centering
\caption{Eight possible combinations of the three crisis scenarios with associated probability of occurrence}
\begin{tabular}{|c|c|c|c||c|}
\hline
\begin{tabular}[c]{@{}c@{}}Event\\   Notation\end{tabular} & Epidemic & Terrorist Attack & Internal Conflict & Prob. Occurrence\\ \hline
None        &               &           &           &8\%     \\ \hline
A           &\cmark         &           &           &2\%        \\ \hline
B           &               &\cmark     &           &8\%        \\ \hline
C           &               &           &\cmark     &32\%                 \\ \hline
AB          &\cmark         &\cmark     &           &2\%        \\ \hline
AC          &\cmark         &           &\cmark     &8\%            \\ \hline
BC          &               &\cmark     &\cmark     &32\%            \\ \hline
ABC         &\cmark         &\cmark     &\cmark     &8\%            \\ \hline
\end{tabular}
\label{tab:crisis combinations}
\end{table}

\subsection{Network Design Changes}
In addition to the demand changes resulting from crisis events, as described above, we expand upon the base model by considering various reasonable supply chain alterations.  We examined eight supply chain variants, incorporating additions such as: capping the number of locations a transshipment node can support, increasing the number of transshipment facilities, allowing flexibility in transshipment node placement, and modifying the objective.  The rationale behind these network variations was to either validate the current network configuration or to recommend modifications for the projected ASCNP to improve its robustness and service levels.  However, since the resulting robustness analysis considered a total of 64 combinations (i.e., the eight supply chain model variants crossed with the eight combinations of crisis scenarios) only the most informative subset of the analysis are included in this manuscript. 

\subsubsection{Transshipment Node Capacity Modification}
In the original network, every transshipment facility location can support all locations.  However, in a real-world setting, such an uncapacitated approach is infeasible.  To address this issue we implemented a modification to ASCNP restricting each transshipment facility to a maximum number of locations they can support.  Based on subject matter expert feedback, we determine the maximum number of demand locations a transshipment hub can support is \textit{C}=26.

\subsubsection{Base Model plus Three Additional Transshipment Nodes}
The base network established seven locations as transshipment nodes.  This model variation, denoted \textit{Base plus Three}, allows the setup of three additional transshipment nodes and identiﬁes their locations.  This could potentially increase the number of connections between locations and transshipment nodes, leading to an improvement in the network service level and a higher resilience.  We incorporate this modification in our base model by replacing Constraint (\ref{Con:NumHubs}) with Constraint (\ref{Con:NumHubsRelaxed}) which ensures that a total of 10 transshipment nodes are selected.

\begin{equation}\label{Con:NumHubsRelaxed}
\begin{aligned}
\quad & \sum_{\textit{i} \in \textit{I}/{1}}{x_i} = 10
\end{aligned}
\end{equation}

\subsubsection{Main Hub with Six Additional Transshipment Nodes}\label{sec:accrawithsix}
This model variation, denoted \textit{Hub with Six}, allows almost complete flexibility in determining the appropriate transshipment facility locations.  The only specified transshipment facility location is the hub – the remaining six transshipment node locations are selected by the model.  This variation provides insight into whether the originally specified locations are appropriate transshipment nodes.  We introduce this variation by replacing Constraint (\ref{Con:StartingHubs}) of the base model with Constraint (\ref{con:accrawithsix}) in which the hub is the only required transshipment node.

\begin{equation}\label{con:accrawithsix}
\begin{aligned}
\quad & {x_{22}} = 1
\end{aligned}
\end{equation}

\subsubsection{Hub with Six Additional Transshipment Nodes and Zero Weight on Connectivity}
This model variant, \textit{Altered Objective}, requires a transshipment node at the hub and allows the model to determine six other transshipment locations (same as \ref{sec:accrawithsix}, however this model additionally alters the objective function.  In previous variations, the distance of shipments had a weight of ${w_1}$=0.95, and the distance of additional connections a weight of ${w_2}$=0.05.  For this variation, we set ${w_1}$=1 and ${w_2}$=0, meaning that the distance of shipments is the sole objective, and the backup (non-shipping) routes could be chosen arbitrarily.  This variation aims to expose the cost such backup connections (resilience) inflicts on the network.  The altered objective variation shows how resilience affects the optimal node and route selection within the model.  %This variant uses the following objective function 13:

%\begin{equation}
%\begin{aligned}
%\quad & \min_{x,y,z} = \sum_{k \in K}\sum_{i \in I}\sum_{j \in I}{D_{i,j}^k}{z_{i,j}^k}
%\end{aligned}
%\end{equation}

In the following section, we analyze these model variants in conjunction with the most frequent demand scenarios, that is scenario C (internal conflict) and scenario BC (terrorist attack and internal conflict).  

\subsection{Model Instances Output}
\subsubsection{Base Model Without Capacity Constraints}
The base model requires the main hub along with six other locations to be transshipment nodes.  Table \ref{tab:base model connections} shows the selected transshipment facilities, the number of locations each facility supplied, and the selected mode of transportation.

\begin{table}[H]
\centering
\caption{Base Model number of connections and transportation mode}
\begin{tabular}{|c|c|c|c|lllll}
\cline{1-4}
Node & Transshipment Facility Location & Number of Connections & Transportation Mode                                     &  &  &  &  &  \\ \cline{1-4}
22   & Main Hub                           & 51                    & \begin{tabular}[c]{@{}c@{}}Air\\   and Sea\end{tabular} &  &  &  &  &  \\ \cline{1-4}
7    & TF-1                     & 4                     & Land                                                    &  &  &  &  &  \\ \cline{1-4}
12   & TF-2                          & 1                     & Land                                                    &  &  &  &  &  \\ \cline{1-4}
19   & TF-3                       & 1                     & Land                                                    &  &  &  &  &  \\ \cline{1-4}
38   & TF-4                          & 1                     & Land                                                    &  &  &  &  &  \\ \cline{1-4}
46   & TF-5                          & 1                     & Sea                                                     &  &  &  &  &  \\ \cline{1-4}
58   & TF-6                           & 0                     & --                                                      &  &  &  &  &  \\ \cline{1-4}
\end{tabular}
\label{tab:base model connections}
\end{table}

%--suggest hiding this figure and the rest of the maps\begin{figure}[H]
  %  \centering
   % \includegraphics[width = 0.85\textwidth]{Fig3.png}
    %\caption{Base Model network flow}
    %\label{fig:base model flow}
%\end{figure}

The main hub is the most used transshipment node in the ASCNP. It supplies 51 locations and leaves only eight other locations to be reached via another transshipment node.  This is primarily due to no available airports at those eight demand locations.  Therefore, it is more efficient to supply these locations by first flying their demand to a transshipment node and then supplying them by land.  Given that there is no limit on the number of locations a transshipment node can supply, the most efficient method in terms of distance is for the hub to ship directly to demand locations instead of going through another transshipment node.  For instance, the distance between TF-1 (a transshipment facility location) and one end demand locations is shorter than the distance between it and the hub.  However, shipping through the TF-1 instead of directly from the hub would increase the shipping distance by approximately 90 miles.

Sea transportation is utilized to ship material from the hub to five locations, each which have seaports relatively near the hub.  When available, sea transportation is the most efficient means of transportation for locations less than 500 miles from the hub.

TF-1 is the second most utilized transshipment node, supplying four locations.  Three transshipment facilities ship to one location by land, while another supplies one location by sea, leaving one current transshipment facility idle--as a transshipment node.  This node is such a large distance from the hub that it could be more suitable as solely a demand location rather than as a transshipment node.  

\subsubsection{Base Model with Capacity Constraints}
In this model instance, we add a capacity constraint to the base model by limiting the number of locations a transshipment node can support to 26.  This modified network begins to reflect the hub and spoke design of the current network that relies on transshipment facilities to reach lower demand end locations.

%\begin{figure}[H]
 %   \centering
  %  \includegraphics[width = 0.85\textwidth]{Fig4.png}
   % \caption{Base Model network flow with transshipment node connections constraint}
    %\label{fig:base model flow constrained}
%\end{figure}

Table \ref{tab:base model tranship} shows the selected transshipment facility locations and the number of locations that each facility supplies.  

\begin{table}[H]
\centering
\caption{Base Model number of transshipment node connections with transshipment node connections constraint}
\begin{tabular}{|c|c|c|llllll}
\cline{1-3}
Node & Transshipment Facility Location & Number of Connections &  &  &  &  &  &  \\ \cline{1-3}
22   & Main Hub                           & 26                    &  &  &  &  &  &  \\ \cline{1-3}
7    & TF-1                     & 12                    &  &  &  &  &  &  \\ \cline{1-3}
46   & TF-2                         & 11                    &  &  &  &  &  &  \\ \cline{1-3}
12   & TF-3                          & 4                     &  &  &  &  &  &  \\ \cline{1-3}
19   & TF-4                       & 4                     &  &  &  &  &  &  \\ \cline{1-3}
38   & TF-5                          & 2                     &  &  &  &  &  &  \\ \cline{1-3}
58   & TF-6                           & 0                     &  &  &  &  &  &  \\ \cline{1-3}
\end{tabular}
\label{tab:base model tranship}
\end{table}

Because of the capacity constraint, more transshipment nodes play a greater role in the shipping distribution than in the base model.  The hub still supplies the most locations (reaching the maximum capacity of 26), but TF-1 and TF-2, being very central in the network, ship to 12 and 11 locations, respectively.  We note that TF-6 is still not used as a distribution center but simply serves as a demand node.  This could indicate the unsuitability of TF-6 as a transshipment node.  

\subsubsection{Base Model with Three Additional Transshipment Nodes with Demand Change }
In this analysis, we study the \textit{Base plus Three} model under the different demand scenarios described in the previous section.  We analyze the demand scenarios for cases of a network both with and without capacity on its transshipment nodes’ connections.
\\
\\
Table \ref{tab:base plus 3 probs} summarizes the likelihood of opening each additional transshipment node based on the realized demand scenarios when no capacity is imposed on the network.  The likelihoods were computed based on the probabilities computed in Table \ref{tab:crisis combinations}.  As the results show, TF-7 and TF-8 open regardless of the realized demand scenario.  As expected, the probability that the third location is established at TF-9 or TF-10 corresponds to the likelihood of the events (B, C, AC) and (None, A, AB, BC, ABC), respectively.

\begin{table}[H]
\centering
\caption{Base plus Three likelihood of transshipment node selection without transshipment node connections constraint}
\begin{tabular}{lllllllll}
\cline{1-4}
\multicolumn{1}{|c|}{Node} & \multicolumn{1}{c|}{Transshipment Facility Location} & \multicolumn{1}{c|}{Likelihood} & \multicolumn{1}{c|}{Scenarios}                &  &  &  &  &  \\ \cline{1-4}
\multicolumn{1}{|c|}{2}    & \multicolumn{1}{c|}{TF-7}                         & \multicolumn{1}{c|}{100\%}      & \multicolumn{1}{c|}{All scenarios}            &  &  &  &  &  \\ \cline{1-4}
\multicolumn{1}{|c|}{42}   & \multicolumn{1}{c|}{TF-8}                           & \multicolumn{1}{c|}{100\%}      & \multicolumn{1}{c|}{All scenarios}            &  &  &  &  &  \\ \cline{1-4}
\multicolumn{1}{|c|}{61}   & \multicolumn{1}{c|}{TF-9}                            & \multicolumn{1}{c|}{51\%}       & \multicolumn{1}{c|}{\{B, C, AC\}}             &  &  &  &  &  \\ \cline{1-4}
\multicolumn{1}{|c|}{20}   & \multicolumn{1}{c|}{TF-10}                         & \multicolumn{1}{c|}{49\%}       & \multicolumn{1}{c|}{\{None, A, AB, BC, ABC\}} &  &  &  &  &  \\ \cline{1-4}
\end{tabular}
\label{tab:base plus 3 probs}
\end{table}

When capacity is imposed on the network, the same additional transshipment facility locations are selected; however their likelihood changes, as summarized in Table \ref{tab:base plus 3 tranship selection prob}.  The transshipment facilities at TF-7 and TF-8 are still opened irrespective of the demand scenario.  The likelihood of using TF-9 as a transshipment node increases to 66\%, and TF-10\textquotesingle s likelihood drops to 34\%.  

\begin{table}[H]
\centering
\caption{Base plus Three likelihood of transshipment node selection with transshipment node connections constraint}
\begin{tabular}{lllllllll}
\cline{1-4}
\multicolumn{1}{|c|}{Node} & \multicolumn{1}{c|}{Transshipment Facility Location} & \multicolumn{1}{c|}{Likelihood} & \multicolumn{1}{c|}{Scenarios}               &  &  &  &  &  \\ \cline{1-4}
\multicolumn{1}{|c|}{2}    & \multicolumn{1}{c|}{TF-7}                         & \multicolumn{1}{c|}{100\%}      & \multicolumn{1}{c|}{All scenarios}           &  &  &  &  &  \\ \cline{1-4}
\multicolumn{1}{|c|}{42}   & \multicolumn{1}{c|}{TF-8}                           & \multicolumn{1}{c|}{100\%}      & \multicolumn{1}{c|}{All scenarios}           &  &  &  &  &  \\ \cline{1-4}
\multicolumn{1}{|c|}{61}   & \multicolumn{1}{c|}{TF-9}                            & \multicolumn{1}{c|}{66\%}       & \multicolumn{1}{c|}{\{None, B, C, AC, ABC\}} &  &  &  &  &  \\ \cline{1-4}
\multicolumn{1}{|c|}{20}   & \multicolumn{1}{c|}{TF-10}                         & \multicolumn{1}{c|}{34\%}       & \multicolumn{1}{c|}{\{A, AB, BC\}}           &  &  &  &  &  \\ \cline{1-4}
\end{tabular}
\label{tab:base plus 3 tranship selection prob}
\end{table}

\subsubsection{Base plus Three with Internal Conflict}
\par In this section, we solve the \textit{Base plus Three model} with a network disruption due to internal conflict (scenario C).  We solve this model instance for cases of a network both with and without the transshipment node connections constraint.  

Table \ref{tab:base plus three no cap} summarizes the transshipment nodes selected as well as their number of established connections for the case of the network instance without connectivity constraints.

\begin{table}[H]
\centering
\caption{Base plus Three number of connections without transshipment node connections constraint }
\begin{tabular}{|c|c|c|llllll}
\cline{1-3}
Node & Transshipment Facility Location & Number of Connections &  &  &  &  &  &  \\ \cline{1-3}
22   & Main Hub                           & 29                    &  &  &  &  &  &  \\ \cline{1-3}
2    & TF-7                         & 15                    &  &  &  &  &  &  \\ \cline{1-3}
61   & TF-9                            & 6                     &  &  &  &  &  &  \\ \cline{1-3}
7    & TF-1                     & 4                     &  &  &  &  &  &  \\ \cline{1-3}
42   & TF-8                           & 2                     &  &  &  &  &  &  \\ \cline{1-3}
12   & TF-3                          & 1                     &  &  &  &  &  &  \\ \cline{1-3}
38   & TF-5                          & 1                     &  &  &  &  &  &  \\ \cline{1-3}
46   & TF-2                          & 1                     &  &  &  &  &  &  \\ \cline{1-3}
19   & TF-4                       & 0                     &  &  &  &  &  &  \\ \cline{1-3}
58   & TF-6                           & 0                     &  &  &  &  &  &  \\ \cline{1-3}
\end{tabular}
\label{tab:base plus three no cap}
\end{table}

The opening of additional transshipment nodes reduces the number of locations supported by the hub to 29.  In fact, with the additional opening of seaport transshipment facilities at TF-9 and TF-7, the hub ships via sea to these locations, who then supply farther locations and the transshipment nodes of TF-2, TF-3 and TF-5, now each support one location.  TF-4 and TF-6 do not supply any locations.

\subsubsection{Main Hub with Six with Demand Change}
In this model variant, the main hub is the only required transshipment node.  We analyze this model modification under the different demand scenarios without the transshipment node connections constraint.  Table \ref{tab:accra with six tranship probs} shows the selected transshipment facility locations and their likelihood of being selected based on the realized demand scenarios.

\begin{table}[H]
\centering
\caption{Hub with Six likelihood of transshipment node selection without transshipment node connections constraint}
\begin{tabular}{lllllllll}
\cline{1-4}
\multicolumn{1}{|c|}{Node} & \multicolumn{1}{c|}{Transshipment Facility Location} & \multicolumn{1}{c|}{Likelihood} & \multicolumn{1}{c|}{Scenarios}            &  &  &  &  &  \\ \cline{1-4}
\multicolumn{1}{|c|}{2}    & \multicolumn{1}{c|}{TF-7}                         & \multicolumn{1}{c|}{100\%}      & \multicolumn{1}{c|}{All}                  &  &  &  &  &  \\ \cline{1-4}
\multicolumn{1}{|c|}{7}    & \multicolumn{1}{c|}{TF-1}                     & \multicolumn{1}{c|}{100\%}      & \multicolumn{1}{c|}{All}                  &  &  &  &  &  \\ \cline{1-4}
\multicolumn{1}{|c|}{20}   & \multicolumn{1}{c|}{TF-10}                         & \multicolumn{1}{c|}{100\%}      & \multicolumn{1}{c|}{All}                  &  &  &  &  &  \\ \cline{1-4}
\multicolumn{1}{|c|}{42}   & \multicolumn{1}{c|}{TF-8}                           & \multicolumn{1}{c|}{100\%}      & \multicolumn{1}{c|}{All}                  &  &  &  &  &  \\ \cline{1-4}
\multicolumn{1}{|c|}{61}   & \multicolumn{1}{c|}{TF-9}                            & \multicolumn{1}{c|}{100\%}      & \multicolumn{1}{c|}{All}                  &  &  &  &  &  \\ \cline{1-4}
\multicolumn{1}{|c|}{38}   & \multicolumn{1}{c|}{TF-5}                          & \multicolumn{1}{c|}{82\%}       & \multicolumn{1}{c|}{\{B, C, AB, AC, BC\}} &  &  &  &  &  \\ \cline{1-4}
\multicolumn{1}{|c|}{17}   & \multicolumn{1}{c|}{TF-11}                           & \multicolumn{1}{c|}{18\%}       & \multicolumn{1}{c|}{\{None, A, ABC\}}     &  &  &  &  &  \\ \cline{1-4}
                   &  &  &  &  & 
\end{tabular}
\label{tab:accra with six tranship probs}
\end{table}

In addition to the hub, transshipment nodes are always selected at five transshipment facilities, irrespective of the scenario.  On the other hand, two locations are used as a transshipment node 82\% and 18\% of the time, respectively.  In comparison with the selected transshipment nodes in the ASCN base model, we note that the only common transshipment facility locations apart from the hub are TF-1 and TF-5.  However, by further inspection we notice that the other transshipment nodes are clustered around those selected by the original decision makers, thus validating the network planners' transshipment nodes' placements.  Though the exact locations are slightly different, we find that the regions are grouped appropriately.

\section{Findings and Contributions}
The AFSCN model selected transshipment nodes to reduce the shipping distance to remote locations.  The locations that only had one form of transportation (mainly land) required a transshipment node between the hubs and the demand locations.  A transshipment facility used the shortest form of transportation initially to ship to other transshipment nodes that then transported the items to the distant demand location. 
\\
\\
Establishing backup connections for resilience leads to solutions that use less efficient routes.  This has a cost, and when decisions are made to account for the distances of backup connections to transshipment nodes, the model does not always choose the shortest shipping distances – the model solves for the shortest distance overall.
\\
\\
The results of the different cases analyzed in the previous sections suggest the following key findings.  First, most of the current transshipment facility locations are appropriately selected.  The transshipment locations were chosen based on operational requirements; the network was not modeled to to optimally select the locations. The six additional transshipment nodes opened in the second variation of AFSCN were either those chosen by the original network designers, or they were in the same geographic area.  Based on current and estimates of future demand, the supply chain planners created a well-designed network. 
\\
\\
Second, one transshipment facility location was never selected in any of the model instances nor under any of the demand scenarios.  This is due to its location in the outermost part of the network, making it distant from most other locations.  This suggests that this location should not be used as a transshipment facility. We recognize however that there are many reasons beyond optimal efficiency and resilience that facility locations are chosen. For example, political or future network design considerations are often taken into account. The value that our model provides to decision makers is that it communicates that under no current or forecasted requirements is the current design optimal. 
\\
\\
Third, in addition to recommending where savings can be achieved by removing a transshipment facility, the results of the \textit{Base plus Three variant}, demonstrate which locations are most suitable for additional transshipment facilities should the future need arise.  Three additional locations are consistently suggested across the different scenarios. With the current AFSCN infrastructure, our method provides recommendations as to how the network should be contracted or expanded.   
\\
\\
A main contribution of this paper is a method by which supply chain designers and network planners can integrate risk assessments, local capabilities and environmental factors into facility location decisions. We provide a risk analysis framework that classifies levels of conflict, infrastructure capability and environmental risk. These factors are generalizable across any facility location problem and are relevant to all supply chains.  We then show how publicly available data sets--provided by the US State Department and CIA Factbook, can be used to populate the risk framework with location-specific assessments. The countries of West Africa and real demand scenarios are used to demonstrate how these factors are taken into account to design an efficient network that is both resilient to disruption and robust to demand changes. To the best of our knowledge, this is the first such effort to do this.

\section{Conclusion and Future Research}
This research study created a resilient, operational road map to allow a network to function during disruptions.  Our AFSCN model incorporates 60 West African demand locations and uses transshipment nodes as originally determined by network planners.  To account for demand fluctuations due to disruption--terrorist attacks, epidemics and/or natural disaster, multiple different scenario occurrences and their likelihood were identified. We then analyzed the resulting network connections and shipping ﬂow.
\\
\\
The first model variation allowed three additional transshipment nodes should they be required.  The second variation of the model allowed the selection of any six transshipment nodes in addition to the main hub, without requiring any specific location to be a transshipment node.  This variation answered whether the chosen locations are efficient for a transshipment facility.  The final variation of the model modiﬁed the objective function by minimizing the distance of the shipping routes without considering the backup connections distances to other warehouses.  This variant assessed the impact of backup connections on transshipment facility locations and established routes.
\\
\\
Four main findings arose from our analysis.  First, the transshipment locations are accurately selected, as each of the transshipment nodes in the Base plus Three and Hub with Six models were either close in proximity or the exact locations as selected by the original network designers.  One location not selected in any of the model variants, no matter which demand scenario occurred.  The fact that this facility is situated in the outermost part of the network makes it unsuitable to be a transshipment node.  Three additional transshipment facilities are identified as the most suitable locations should the need arise.  These locations are seaports along the coast of Ghana.  
\\
\\
Finally, establishing backup connections for resilience leads to solutions that use less efficient routes.  In fact, when accounting for the distances of backup connections to transshipment nodes, the model does not always choose the shortest shipping distance but solves for the shortest distance overall. Two key points emerge from this insight. First, there is an efficiency cost to resilience. Modeling the network allows us to determine that cost. Second, resilience must be considered at the supply chain level and not the individual node level. As our AFSCN model demonstrates, the optimal system design may lead to sub-optimal solutions for individual nodes and routes. 
\\
\\
This study opens several opportunities for future research.  In the current model, there is no cost imposed for creating a transshipment node or for the shipping distance.  The objective of the model could be altered to account for these financial aspects, as they may play a role in the actual determination of transshipment facility placement.  Further, the number of items shipped by air, land, and sea is unconstrained.  However, in real-world situations, different locations have different capacity and means available, e.g., air crew sizes, number of land/sea vehicles, etc.  A more realistic instance of this model would be to impose a limit on the quantity of material that can shipped by each mode.  Finally, in this work, the demand uncertainty is treated by individually assessing the network modification that occurs following the realization of the demand scenario events with the highest likelihood.  A different approach would be to incorporate the demand randomness in the model and solve for the resulting transshipment node placements, established routes and links, and network flow quantities through stochastic or robust optimization.  Undoubtedly, such an approach would require significant nontrivial modifications to our model.   

%\section*{Disclaimer} The views expressed in this article are those of the authors and do not reflect the official policy or position of the United States Air Force, the Department of Defense, or the United States Government.

\bibliography{WALNbibfile}

\end{document}